\documentstyle[11pt,aaspp4,flushrt,tighten]{article}
\singlespace
\def\HI{\hbox{H~$\scriptstyle\rm I\ $}}
\def\HII{\hbox{H~$\scriptstyle\rm II\ $}}
\def\nHI{{\rm HI}}
\def\nH{{\rm H}}
\def\nHII{{\rm HII}}
\def\nHe{{\rm He}}

\def\nHeII{{\rm HeII}}
\def\nHeIII{{\rm HeIII}}
\def\bHII{\hbox{\bf H~$\scriptstyle\bf II\ $}}
\def\HeI{\hbox{He~$\scriptstyle\rm I\ $}}
\def\HeII{\hbox{He~$\scriptstyle\rm II\ $}}
\def\bHeII{\hbox{\bf He~$\scriptstyle\bf II\ $}}
\def\HeIII{\hbox{He~$\scriptstyle\rm III\ $}}
\def\CIV{\hbox{C~$\scriptstyle\rm IV\ $}}

\def\SiIV{\hbox{Si~$\scriptstyle\rm IV\ $}}
\def\kms{\,{\rm km\,s^{-1}}}
\def\kmsmpc{\,{\rm km\,s^{-1}\,Mpc^{-1}}}
\def\cmm{\,{\rm cm^{-2}}}

\def\uvunits{\,{\rm ergs\,cm^{-2}\,s^{-1}\,Hz^{-1}\,sr^{-1}}}
\def\emunits{\,{\rm ergs\,s^{-1}\,Hz^{-1}\,Mpc^{-3}}}
\def\ndotunits{\,{\rm s^{-1}\,Mpc^{-3}}}
\def\ndotun{\,{\rm phot\,s^{-1}}}
\def\msun{\,{\rm M_\odot}}

\def\sfrd{\,{\rm M_\odot\,yr^{-1}\,Mpc^{-3}}}
\def\sfr{\,{\rm M_\odot\,yr^{-1}}}

\def\Lya{Ly$\alpha\ $}
\def\etal{{et al.\ }}
\def\AB{{\rm AB}}

\def\spose#1{\hbox to 0pt{#1\hss}}
\def\lta{\mathrel{\spose{\lower 3pt\hbox{$\mathchar"218$}}
     \raise 2.0pt\hbox{$\mathchar"13C$}}}
\def\gta{\mathrel{\spose{\lower 3pt\hbox{$\mathchar"218$}}
     \raise 2.0pt\hbox{$\mathchar"13E$}}}

\begin{document}
\title{Radiative Transfer in a Clumpy Universe: III. The Nature of Cosmological
Ionizing Sources}

\author{Piero Madau} 
\affil{Space Telescope Science Institute, 3700 San Martin 
Drive, Baltimore MD 21218}

\author{Francesco Haardt}
\affil{Dipartimento di Fisica, Universit\'a degli Studi di Milano, via Celoria
16, 20133 Milano, Italy} 

\and
\author{Martin J. Rees}
\affil{Institute of Astronomy, Madingley Road, Cambridge CB3 0HA, UK}

\begin{abstract}

The history of the transition from a neutral intergalactic medium (IGM) to one
that is almost fully ionized can reveal the character of cosmological ionizing
sources. We study the evolution of the volume filling factors of \HII and \HeIII
regions in a clumpy IGM, and discuss the implications for rival reionization 
scenarios of the rapid decline in the space density of radio-loud quasars and
of the large population of star-forming galaxies recently observed at $z\gta 
3$. The hydrogen 
component in a highly inhomogeneous universe is completely reionized when the 
number of photons emitted above 1 ryd in one recombination time equals the 
mean number of hydrogen atoms. If stellar sources are responsible for keeping 
the IGM ionized at $z=5$, the rate of star formation at this epoch must 
be comparable or greater than the one inferred from optical observations of 
galaxies at $z\approx 3$, and the mean metallicity per baryon in the universe 
$\sim 1/500$ solar. An early generation of stars in dark matter halos with 
circular velocities $v_{\rm circ}\approx 50\,\kms$, possibly one of the main 
source of UV photons at high-$z$, could be detectable with the 
{\it Next Generation Space Telescope}. Models in which the quasar emissivity
declines rapidly at $z\gta3$ predict a late \HeII reionization epoch, a feature
that could explain the recent detection of patchy \HeII \Lya at $z=2.9$ by 
Reimers \etal (1997) and the abrupt change observed by Songaila (1998) at 
about the same epoch of the \SiIV/\CIV ratio, but appear unable to provide the 
required number of hydrogen-ionizing photons at $z\approx 5$. 

\end{abstract}


\section{Introduction}

The existence of a filamentary, low-density intergalactic medium (IGM), which
contains the bulk of the hydrogen and helium in the universe, is predicted as a
product of primordial nucleosynthesis (e.g., Copi, Schramm, \& Turner 1995) and
of hierarchical models of gravitational instability with ``cold dark matter''
(CDM) (Cen \etal 1994; Zhang, Anninos, \& Norman 1995; Hernquist \etal 1996).
The application of the Gunn-Peterson (1965) constraint on the amount of smoothly
distributed neutral material along the line of sight to distant objects 
requires the hydrogen component of the diffuse IGM to have been highly ionized 
by $z\approx 5$ (Schneider, Schmidt, \& Gunn 1991), and the helium component 
by $z\approx 2.5$ (Davidsen, Kriss, \& Zheng 1996). The plethora of discrete 
absorption systems which give origin to the \Lya forest in the spectra of 
background quasars are also inferred to be strongly photoionized. It thus 
appears that substantial sources of ultraviolet
photons were present at $z>5$, perhaps low-luminosity quasars (Haiman \&
Loeb 1998) or a first generation of stars in virialized dark matter halos with
$T_{\rm vir}\gta 10^4\,$K (Couchman \& Rees 1986; Tegmark, Silk, \& Blanchard
1994; Fukugita \& Kawasaki 1994; Ostriker \& Gnedin 1996; Ciardi \& Ferrara
1997; Haiman \& Loeb 1997; Miralda-Escud\`e \& Rees 1998). Early star
formation provides a possible explanation for the widespread existence of heavy
elements in the IGM (Cowie \etal 1995), while reionization by QSOs may
produce a detectable signal in the radio extragalactic background at meter
wavelengths (Madau, Meiksin, \& Rees 1997). Establishing the character of
cosmological ionizing sources is an efficient way to constrain competing
models for structure formation in the universe, and to study the collapse and
cooling of small mass objects at early epochs. While the nature, spectrum, and 
intensity of the background UV flux which is responsible for maintaining the 
intergalactic gas and the Ly$\alpha$ clouds in a highly ionized state at $z\lta
3$ has been the subject of much debate in the last decade (e.g., Bechtold 
\etal 1987; Shapiro \& Giroux 1987; Songaila, Cowie, \& Lilly 1990; 
Miralda-Escud\`e \& Ostriker 1990; Madau 1992; Sciama 1995; Haardt \& Madau 
1996; Giroux \& Shapiro 1996), it is only in the past few years that new 
observations have provided reliable information on the presence and physical
properties of sources and sinks (due to continuum opacities) of UV radiation 
in the interval $3\lta z\lta 5$. 

This is the third paper in a series aimed at a detailed study of the 
production, absorption, and reprocessing of ionizing photons in a clumpy 
universe. In Paper I (Madau 1995) we showed how the stochastic attenuation 
produced by neutral
hydrogen along the line of sight can seriously distort our view of objects at 
cosmological distances. In Paper II (Haardt \& Madau 1996) a numerical code was
developed which follows the propagation of photoionizing radiation through 
an IGM in which the \HI and \HeII region networks have fully percolated at 
the redshifts of interest. It is the purpose 
of this paper to focus instead on the candidate sources of photoionization at 
early times and on the time-dependent reionization problem, i.e. on the history
of the transition from a neutral IGM to one that is almost fully ionized. The 
primary motivation of this study can be found perhaps in the simple realization
that the {\it breakthrough epoch} (when all radiation sources can see 
each other in the Lyman continuum) occurs much later in the universe than 
the {\it overlap epoch} (when individual ionized zones become simply 
connected and every point in space is exposed to ionizing radiation), and that
at high redshifts the ionization equilibrium is actually determined by
the {\it instantaneous } UV production rate. The fact that the IGM is still
optically thick at overlapping, coupled to recent 
observations of a rapid decline in the space density of radio-loud quasars 
and of a large population of star-forming galaxies at $z\gta 3$, will 
have some interesting implications for rival ionization scenarios.

Throughout this paper we will adopt an Einstein-de Sitter universe ($q_0=0.5$) 
with $H_0=50h_{50}\, \kmsmpc$.

\section{The Universe After Complete Overlapping}

The complete reionization of the universe manifests itself in the absence of a
Gunn-Peterson absorption trough in the spectra of galaxies and quasars at 
$z\lta 5$. In the presence of a uniform medium of \HI density $n_\nHI(z)$ 
along the path to a distant object, the optical depth associated with 
resonant \Lya scattering at $\lambda_\alpha(1+z)$ is 
\begin{equation}
\tau_{\rm GP}(z)={\pi e^2 f_\alpha \lambda_\alpha n_\nHI(z)\over m_ec H}=
8.3\times 10^{10} h_{50}^{-1} {n_\nHI(z)\over (1+z)^{3/2}}, 
\label{eq:gp} 
\end{equation}
where $H$ is the Hubble constant, $f_\alpha$ is the oscillator strength of 
the transition, and all other symbols have their usual meaning. The same 
expression for the opacity is also valid in the case of optically thin (to
\Lya scattering) discrete clouds as long as $n_\nHI$ is replaced with
the average neutral density of individual clouds times their volume filling
factor. The strongest 
limit on the amount of diffuse intergalactic neutral hydrogen at $z=4.3$ is 
provided by a direct measurement of the quasar Q1202--0725 flux in regions of 
the spectrum where lines are absent (Giallongo \etal 1994): a best estimate of 
$\tau_{\rm GP}=0.02\pm 0.03$ is found. Even if 99\% of all the cosmic baryons 
fragment at these epochs into structures that can be identified with QSO 
absorption systems, with only 1\% remaining in a smoothly distributed 
component (cf. Zhang \etal 1998), the implication is a diffuse IGM which is 
ionized to better than 1 part in $10^4$. Except at the highest column densities,
discrete absorbers are also inferred to be strongly photoionized. From QSO 
absorption studies we also know that neutral hydrogen accounts for only a 
small fraction, $\sim 10\%$, of the nucleosynthetic baryons at early epochs
(e.g. Lanzetta, Wolfe, \& Turnshek 1995). 

In this section we shall discuss the last stage of the reionization process,
when individual ionized zones have overlapped, the reionization of hydrogen 
(and helium) in the universe has been completed, and the IGM has been exposed
everywhere to Lyman continuum photons. An often overlooked point is worth 
remarking here, namely the fact that even if the bulk of the baryons in the 
universe are fairly well ionized at $z\lta 5$, the residual neutral hydrogen 
still present in the Ly$\alpha$ forest clouds and the Lyman-limit 
systems (LLS) along the line of sight significantly attenuates the ionizing 
flux from cosmologically distant sources. In other words, while the complete
overlapping of I-fronts occurs at $z\gta 5$, the universe does not 
become optically thin to Lyman continuum photons until much later, at a 
``breakthrough'' redshift $z_{\rm br}\approx 1.6$ (see \S~2.3). It is only 
after breakthrough that every point in space at any redshift $z<z_{\rm br}$ 
will be exposed to ionizing photons from {\it all} radiation sources out 
to $z_{\rm br}$.

\subsection{Cosmological Radiative Transfer}

The radiative transfer equation in cosmology describes the time evolution of
the specific intensity $J$ of a diffuse radiation field (e.g., Peebles 1993): 
\begin{equation}
\left({\partial \over \partial t}-\nu {\dot a \over a} {\partial
\over \partial \nu}\right)J=-3{\dot a \over a}J-c\kappa J + 
{c\over 4\pi}\epsilon, \label{eq:rad} 
\end{equation}
where $a$ is the scale parameter, $c$ the speed of the light,
$\kappa$ the continuum absorption coefficient per unit length along the line of
sight, and $\epsilon$ is the proper space-averaged volume emissivity. The mean
specific intensity of the radiation background at the observed frequency 
$\nu_o$, as seen by an observer at redshift $z_o$, is then 
\begin{equation}
J(\nu_o,z_o)={1\over 4\pi}\int_{z_o}^{\infty}\, dz\, {dl \over dz} 
{(1+z_o)^3 \over (1+z)^3} \epsilon(\nu,z)e^{-\tau_{\rm eff}}, \label{eq:Jnu} 
\end{equation}
where $\nu=\nu_o(1+z)/(1+z_o)$, and $dl/dz$ is the line element in a Friedmann
cosmology. The effective optical depth $\tau_{\rm eff}$ due to discrete 
absorption systems is defined, for Poisson-distributed clouds, as
\begin{equation}
\tau_{\rm eff}(\nu_o,z_o,z)=\int_{z_o}^z\, dz'\int_0^{\infty}\, dN_\nHI\,
{\partial^2 N \over \partial N_\nHI \partial z'} (1-e^{-\tau}) \label{eq:tau} 
\end{equation}
(Paresce, McKee, \& Bowyer 1980), where $\partial^2 N / \partial N_\nHI\partial
z'$ is the redshift and column density distribution of absorbers along the line
of sight, and $\tau$ is the Lyman continuum optical depth through an individual
cloud. 

\subsection{Intervening Absorption}

The actual distribution of intervening clouds
is still quite uncertain, especially in the range $10^{16.3}\lta N_\nHI\lta
10^{17.3}\cmm$, where most of the contribution to the effective photoelectric
opacity actually occurs. As a function of \HI column, a single power-law with
slope $-1.5$  appears to provide a surprisingly good description over
nearly 10 decades in $N_\nHI$ (Hu \etal 1995; Petitjean \etal 1993; Tytler
1987). In a recent analysis of the density distribution 
by redshift interval, Kim \etal (1997) have used high resolution, high S/N {\it
Keck} spectra to show that the deficit of lines (relative to this fit) noted by
Petitjean \etal above $N_\nHI>10^{14.3} \cmm$,
while more pronounced at lower redshifts, tends to disappear (or moves to
columns higher than $10^{16}\cmm$) at $z\gta 3.5$. At high redshift, it is
then a good approximation to use for the distribution of absorbers along the 
line of sight: 
\begin{equation}
{{\partial^2N}\over{\partial N_\nHI\partial z}}=N_0\,N_\nHI^{-1.5}(1+z)^
{\gamma}. \label{eq:dis} 
\end{equation}
While it has been assumed in the past that the Ly$\alpha$ forest clouds and LLS
evolve at different rates (hence that there is an actual mismatch around
$10^{17}\cmm$ in the adopted function), recent determinations have derived
a value of $\gamma=1.55\pm 0.39$ for the LLS (Stengler-Larrea \etal 1995),
consistent within the errors with $\gamma=2.46\pm 0.37$ obtained 
by Press \& Rybicki
(1993), $\gamma=1.89\pm 0.28$ by Bechtold (1994), and $\gamma=2.78\pm 0.71$ by
Kim \etal (1997) for the Ly$\alpha$ forest line. For simplicity, we will 
assume here a single redshift exponent, $\gamma=2$, for the entire range in
column densities. A normalization value of $N_0=4.0\times 10^7$ produces about
3 LLS per unit redshift at $z=3$, as observed by Stengler-Larrea \etal (1995),
and, at the same epoch, $\sim 150$ lines above
$N_\nHI=10^{13.77} \cmm$, in good agreement with the estimates of Kim \etal
(1997). With this normalization and $\gamma=2$, the adopted distribution 
provides about the same \HI photoelectic opacity as in the model discussed in
Paper II, but underpredicts the attenuation of quasar continua due to \Lya 
line blanketing (cf Paper I). 

\subsection{Attenuation Length} 

If we extrapolate the $N_\nHI^{-1.5}$ power-law in equation (\ref{eq:dis}) to
very small and large columns, the effective optical depth becomes an analytical
function of redshift and wavelength, 
\begin{equation}
\tau_{\rm eff}(\nu_o,z_o,z)={4\over 3}\sqrt{\pi\sigma_0}\, N_0 \left({\nu_o
\over \nu_L}\right)^{-1.5} (1+z_o)^{1.5}\left[(1+z)^{1.5}-(1+z_o)^{1.5}\right], 
\label{eq:t} 
\end{equation}
where $\sigma_0$ is the hydrogen photoionization cross-section at the Lyman
edge $\nu_L$, and we have not included the contribution of helium to the
attenuation along the line of sight. [While the opacity due to \HeI is 
negligible in the case of a QSO-dominated background, \HeII absorption on the
way must be included for $\nu_o>4\nu_L (1+z_o)/(1+z)$.]

It is practical, for the present discussion, to define a redshift $z_1(\nu_1)$
such that the effective optical depth between $z_o$ and $z_1$ is unity. When 
$\nu_o<\nu_L$, a photon emitted at $z_1$ with frequency $\nu_1=\nu_o (1+z_1)/
(1+z_o)$ will be redshifted below threshold before being significantly 
attenuated by intervening \HI. From equation (\ref{eq:t}) it can be shown that 
the universe will be optically thin below   
\begin{equation}
z_1(\nu_1)={1.616\, \nu_1/\nu_L\over [(\nu_1/\nu_L)^{3/2}-1]^{1/3}}-1.
\end{equation}
This expression gives $z_1=1.85$ for $\nu_1=1.2\nu_L$ and $z_1=1.64$ for
$\nu_1=2\nu_L$. It has a minimum for $\nu_1=1.59\nu_L$, corresponding to what
we shall term in the following the ``breakthrough epoch'' of the universe,
$z_{\rm br}\equiv z_1(1.59\nu_L)=1.56$. For $z<z_{\rm br}$,  all radiation 
sources will be able to see each other in the hydrogen Lyman continuum. Hence, 
it is only at $z<z_{\rm br}$ that the degree of ionization of the IGM will be 
determined by the balance between radiative recombinations and the 
total ionizing flux emitted (say) by all QSOs which appear after $z_{\rm br}$. 

By contrast, due to the rapid increase with lookback time of the number of
absorbers, the mean free path of photons at $912\,$\AA\ becomes so small 
beyond a redshift of 2 that the radiation is 
largely ``local'', as sources at higher redshifts are severely attenuated.  
Expanding equation (\ref{eq:t}) around $z=3$, for example, one gets $\tau_{\rm 
eff} (\nu_L)\approx 0.36 (1+z)^2 \Delta z=1$ for $\Delta z=0.18$. This 
corresponds to a proper distance or ``absorption length'' of only 
\begin{equation}
\Delta l(\nu_L)\approx 33\, {\rm Mpc} \left({1+z\over 4}\right)^{-4.5},
\label{eq:dll} 
\end{equation}
shorter than the mean free path between the rarer LLS because of the additional
absorption from the numerous \Lya forest clouds.\footnote{As filtering through 
a clumpy IGM significantly steepens the UV background spectrum 
(Miralda-Escud\`e \& Ostriker 1990; Madau 1992), the absorption length at 
the \HeII edge is smaller than at 1 ryd, $\Delta l(4\nu_L)\approx 1.5 \Delta l
(\nu_L) \sqrt{J_{228}/J_{912}}$.}\ The small absorption length is mostly due 
to systems with continuum optical depth around unity. In the local (or 
``source-function'') solution to
the equation of radiative transfer, this strong attenuation effect can be 
approximated by simply ignoring sources with $z>z_1$ and neglecting absorption
for those sources with $z<z_1$.  Since only radiation sources within the
volume defined by an absorption length contribute then to the background 
intensity shortward of the Lyman edge, cosmological effects such as source 
evolution and frequency shifts can be neglected, and it is possible to write 
$4\pi J(\nu)\approx \epsilon(\nu)/ \kappa(\nu)=\epsilon(\nu) 
\Delta l(\nu)$. In this approximation the number of ionizing photons per unit 
proper volume present at redshift $z$ is given by 
\begin{equation}
n_{\rm ion}\equiv {4\pi\over c} \int_{\nu_L}^\infty d\nu {J(\nu)\over h\nu}=
\dot n_{\rm ion} {\langle \Delta l \rangle\over c},
\label{eq:nion}
\end{equation}
where $\dot n_{\rm ion}(t)\equiv\int_{\nu_L}^\infty d\nu \epsilon(\nu,t)/h\nu$, 
and $\langle \Delta l \rangle$ is an average over the incident photon 
spectrum.\footnote{Since $\Delta l(\nu)\propto (\nu/\nu_L)^{1.5}$, one
has $\langle \Delta l\rangle=\Delta l(\nu_L) \delta (4^{1.5-\alpha_s}-1)/
(1-4^{-\alpha_s})$, where $\delta\equiv \alpha_s/(1.5-\alpha_s)$,
$\epsilon(\nu)\propto \nu^{-\alpha_s}$, and we have assumed a cutoff at 4 ryd
because of \HeII absorption. A spectrum with $\alpha_s=1.8$ yields 
$\langle \Delta l\rangle=2.2\Delta l(\nu_L)$.}\ From the equation of ionization
equilibrium, $\dot n_{\rm ion}=\bar{n}_\nH/\bar{t}_{\rm rec}$, an approximate
relation can be derived between the absorption length and the volume-averaged 
gas recombination time $\bar{t}_{\rm rec}$ if all absorbers are optically 
thin, highly ionized in both H and He, and contain most of the baryons of the 
universe,
\begin{equation}
\langle \Delta l\rangle\approx {n_{\rm ion}\over \bar{n}_\nH}\, c\bar{t}_{\rm 
rec}, \label{eq:dl} 
\end{equation}
where $\bar{n}_\nH$ is the mean hydrogen density of the expanding IGM, 
$\bar{n}_\nH(0)=1.7\times 10^{-7}$ $(\Omega_b h_{50}^2/0.08)$,
\begin{equation}
\bar{t}_{\rm rec}=[(1+2\chi) \bar{n}_\nH \alpha_B\,C]^{-1}=0.3\, {\rm Gyr} 
\left({\Omega_b h_{50}^2 \over 0.08}\right)^{-1}\left({1+z\over 4}\right)^{-3} 
C_{30}^{-1}, 
\end{equation}
$\alpha_B$ is the recombination coefficient to the excited states of hydrogen,
$\chi$ the helium to hydrogen cosmic abundance ratio, $C\equiv \langle
n_\nHII^2\rangle/\bar{n}_\nHII^2$ is the ionized hydrogen clumping factor,
\footnote{This may be somewhat lower than the total gas clumping factor if 
higher density regions are less ionized (Gnedin \& Ostriker 1997).}\ and 
a gas temperature of $10^4\,$K has been assumed. Clumps which are dense
and thick enough to be self-shielded from UV radiation will stay neutral and
will not contribute to the recombination rate. An empirical determination of 
the  clumpiness of the IGM at high redshifts is hampered by our poor 
knowledge of the ionizing background intensity and the typical size and 
geometry of the absorbers. Numerical N-body/hydrodynamics simulations of 
structure formation in the IGM within the framework of CDM dominated 
cosmologies (Cen \etal 1994; Zhang \etal 1995; Hernquist \etal 1996) have 
recently provided a definite picture for the origin of intervening absorption 
systems, one of an interconnected network of sheets and filaments, with 
virialized systems located at their points of intersection.  In the simulations
of Gnedin \& Ostriker (1997), for example, the clumping factor rises above 
unity when the collapsed fraction of baryons becomes non negligible, i.e. $z\lta 
20$, and grows to $C\gta 10$ (40) at $z\approx 8$ (5) (because of finite 
resolution effects, numerical simulations will actually underestimate 
clumping): the recombination timescale is much shorter than that for a uniform 
IGM, and always shorter than the expansion time.

\subsection{Sources of Ionizing Radiation at High Redshifts}

\subsubsection{The Quasar Cutoff}

The existence of a decline in the space density of bright quasars at redshifts
beyond $\sim 3$ was first suggested by Osmer (1982), and has been since then
the subject of a long-standing debate. In recent years, several optical surveys
have consistently provided new evidence for a turnover in the QSO counts.
The six-band multicolor survey of Warren, Hewett, \& Osmer (1994, hereafter
WHO) shows a drop by a factor of $e^{-3.33(z-3.3)}$ beyond $z\approx 3.3$.
A decline by a factor of 2.7 per unit redshift in the range $2.75<z<4.75$ is
seen for $M_B<-26$ by the Schmidt, Schneider, \& Gunn (1995) grism survey.
Kennefick, Djorgovski, \& de Carvalho (1995, hereafter KDC) found a decrease in
the space density of QSOs with $M_B<-27$ by a factor of $\sim 7$ from $z\approx
2$ to $z\approx 4.35$, and no evidence for any luminosity dependent evolution. 
According to Hawkins \& Veron (1996), the space density relation is rather 
flat in the range
$1.5<z<3.0$, and then gradually decline to $z\approx 4.5$ for $M_R<-25.5$. The
interpretation of the drop-off observed in optically selected samples is
equivocal, however, because of the possible bias introduced by dust obscuration
arising from intervening systems (Ostriker \& Heisler 1984; Wright 1990; Fall
\& Pei 1993). Radio emission, on the other hand, is unaffected by dust, and
Shaver \etal (1996) have recently shown that the space density of radio-loud
quasars also decreases strongly for $z>3$, demonstrating that the turnover is
indeed real and that dust along the line of sight has a minimal effect on
optically-selected QSOs. 

The comoving space density of bright quasars as a function of redshift is
plotted in Figure 1. The data points are taken from Hartwick \& Shade (1990)
(a compilation of 15 major optical surveys with more than 1000 quasars
covering the range $0.1\lta z\lta 3.3$), WHO (who have combined their own and
four other surveys to form a sample which contains about 200 quasars at $2\lta
z\lta 4.5$), Schmidt \etal (1995) (90 quasars in the interval $2.75\lta z\lta
4.75$), and KDC (10 quasars at $z>4$). All of them refer to QSOs with
$M_B<-26$ except for KDC, whose survey limit is one magnitue brighter.
The points have been
normalized to the $z=2.5$ space density of $M_B<-26$ quasars ($M_B<-27$ in the
case of KDC) as given by WHO. For comparison, similarly normalized space
densities are shown for the Parkes flat-spectrum radio-loud sample of Hook,
Shaver, \& McMahon (1998). 

A successful analytical representations for the shape of the quasar blue
luminosity function (LF) is the standard double power-law (Boyle, Shanks, \&
Peterson 1988; Pei 1995): 
\begin{equation}
\phi(L,z)={\phi_*/L_*(z) \over [L/L_*(z)]^{\beta_1}+[L/L_*(z)]^{\beta_2}}.
\label{eq:phi}  
\end{equation}
The data in Figure 1 can be well fit by a simple model in which the entire LF
shifts along the luminosity axis as the position of the break $L_*$ evolves
with redshift,
\begin{equation}
L_*(z)=L_*(0)(1+z)^{\alpha_s-1}~{e^{\zeta z}(1+e^{\xi z_*})\over e^{\xi z}
+ e^{\xi z_*}}, \label{eq:lz}  
\end{equation}
where a power-law spectral distribution for the typical quasar spectrum has
been assumed, $L(\nu)\propto \nu^{-\alpha_s}$, and the dependence on the
spectral index $\alpha_s$ is explicitly shown. The comoving density of quasars
brighter than an absolute magnitude limit $M_{\rm lim}$ as a function
of redshift is given by 
\begin{equation}
N(z,M_B<M_{\rm lim})=\int_{-\infty}^{M_{\rm lim}}dM_B \phi(M_B,z). 
\label{eq:NQ}  
\end{equation}
In a cosmology with $(h_{50}, q_0)=(1, 0.5)$, the LF fitting parameters values
given by Pei (1995) are $\log\ (\phi_*/{\rm Gpc^{-3}})=2.95$, $\beta_1=1.64$,
$\beta_2=3.52$, $M_*(0)=-22.35$. The combination of exponentials we choose in
this paper (to be compared with Pei's representation, a simple Gaussian form
that appears to underestimate the number of quasars at $z\sim 4.5$) to
describe the evolution of quasars over the whole range of redshifts has fitting
parameters $z_*=1.9$, $\zeta=2.58$, and $\xi=3.16$. The results of the model
are plotted in Figure 1 for $\alpha_s=0.5$. The density of QSOs has a 
relatively flat maximum at $1.8\lta z\lta 2.8$, and declines gradually 
at higher redshits. Note that, while at $z\lta 2.5$ and $z\gta 4$ all
estimates are in good agreement with each other, in the range $2.5<z<3.5$ the
Hartwick \& Shade and Schmidt \etal data points appear to be sistematically
lower than the WHO point, a discrepancy that may reflects the
uncertainties arising in the incompleteness corrections, different selection
criteria, etc. 

\subsubsection{Quasar Emissivity}

The QSO emission rate of hydrogen ionizing photons per unit comoving volume 
can then be written as
\begin{equation}
\dot{\cal N}_Q(z)=(1+z)^{\alpha_s-1}~{e^{\zeta z}(1+e^{\xi z_*})\over e^{\xi z}
+ e^{\xi z_*}} {\epsilon_Q(\nu_B,0)\over \alpha_s h} (1-4^{-\alpha_s})
\left({\nu_L\over \nu_B}\right)^{-\alpha_s},   \label{eq:NdQ}
\end{equation}
where $\epsilon_Q(\nu_B,0)$ is the extrapolated $z=0$ emissivity in the
blue band,
\begin{equation}
\epsilon_Q(\nu_B,0)=\int_{L_{\rm min}}^{\infty}\phi(L,0)LdL\approx 9.6\times 
10^{23}\emunits, 
\end{equation}
and, as usual, we have assumed a cutoff in the spectrum at 4 ryd because of 
\HeII absorption. The LF of Seyfert galaxies matches remarkably well that of
optically selected QSOs at $M_B=-23$, and does not show clear evidence of
leveling off down to $M_B\approx -18$ (Cheng \etal 1985). We therefore
integrate the LF from $L_{\rm min}\approx 0.018L_*(0)$, keeping in mind that,
for the adopted shape of the LF, a value of $L_{\rm min}$ 2 times smaller would
only increase the quasar emissivity by about 5\%. On the other hand, a steeper
LF with $\beta_1=2.0$ and same $L_{\rm min}$ would boost $\epsilon_Q(\nu_B,0)$
up by a factor 1.75.

Equation (\ref{eq:NdQ}) assumes a single power-law from optical to EUV
frequencies, a poor approximation to the ``typical'' quasar spectrum. In the
following we shall adopt, except when stated otherwise, a more realistic form
for the quasar spectral energy distribution: 
\begin{equation}
L({\nu})\propto \cases{\nu^{-0.3} &($2500<\lambda<4400\,$\AA);\cr
\noalign{\vskip3pt}\nu^{-0.8} &($1050<\lambda<2500\,$\AA);\cr
\noalign{\vskip3pt}\nu^{-1.8} &($\lambda<1050\,$\AA),\cr}
\end{equation}
where the different slopes have been continuously matched. This is based on the
rest-frame optical and UV spectra of Francis \etal (1991) and Sargent, Steidel,
\& Boksenberg (1989), and is similar to the model used in Paper II except for
a steeper index at the shortest wavelengths, a choice more consistent with 
the recent {\it Hubble Space Telescope} (HST) observations of the EUV spectra 
of radio-quiet QSOs at intermediate redshifts (Zheng \etal 1997; see also 
Laor \etal 1997). 

In Figure 2 we plot the value of $\dot{\cal N}_Q$ as a function of 
redshift according to what will become in the following our ``standard 
QSO model".  It is important
to notice that the procedure adopted to derive this quantity implies a large
correction for incompleteness at high-$z$. With a fit to the quasar
LF which goes as $\phi(L)\propto L^{-1.64}$ at the faint end, the
contribution to the emissivity converges rather slowly, as $L^{0.36}$. At
$z=4$, for example, the blue magnitude at the break of the LF is $M_*\approx 
-25.4$, comparable or slightly fainter than the limit of current high-$z$  
QSO surveys. A large fraction, about 90\% at $z=4$ and even higher at 
earlier epochs, of the ionizing emissivity in our model is therefore 
produced by quasars that have not been actually observed, and are
assumed to be present based on an extrapolation from lower redshifts. While the
value of $\dot{\cal N}_Q$ obtained by including the contribution from 
{\it observed} quasars only would be much smaller at high redshifts than 
shown in Figure 2, it is also fair to ask whether an excess of low-luminosity 
QSOs, relative to the best-fit LF, could actually boost the estimated ionizing 
emissivity at early epochs. The interest in models where the quasar LF 
significantly steepens with lookback time, and therefore predict many more 
QSOs at faint magnitudes than the extrapolation of Pei's (1995) fitting 
formulae, stems from recent claims
of a strong linear correlation between bulge and observed black hole masses
(Magorrian \etal 1997), linked to the steep mass function of dark matter haloes
predicted by hierarchical cosmogonies (Haehnelt \& Rees 1993; Haiman \& Loeb
1998; Haehnelt, Natarajan, \& Rees 1998). We shall see in the next section how
the space density of low-luminosity quasars at high-$z$ may be constrained by 
the observed lack of red, unresolved faint objects in the {\it Hubble Deep 
Field} (HDF). 

\subsubsection{Quasar Candidates in the Hubble Deep Field}

Several surveys of unresolved objects in the HDF have appeared in the 
literature (Elson, Santiago, \& Gilmore 1996; Flynn, Gould, \& Bahcall 1996;
M\'{e}ndez \etal 1996). While the small size of the field (about 5.3 square
arcmin covered by the three WFC chips) does not allow one to set meaningful 
constraints to the number density of bright QSOs, its depth and resolution can 
be used to explore quasars counts to very faint magnitudes (Conti \etal 1998). 
In the following we shall denote with $B_\AB, V_\AB$, and $I_\AB$ the AB 
magnitudes in the F450W, F606W, and F814W HDF bandpasses.

Elson \etal find 59 unresolved objects down to a limiting magnitude of
$I_\AB\le 27.2$.\footnote{Throughout this paper we shall denote with 
$B_\AB, V_\AB$, and $I_\AB$ the AB magnitudes in the F450W, F606W, and F814W 
HDF bandpasses.}\ The bulk of them are faint and blue, $(V-I)_\AB\le 1.6$ mag,
with only the 9 brighter sources scattering out to $(V-I)_\AB\le3.1$. The number
of point-like or nearly so objects identified by M\'{e}ndez \etal and Flynn 
\etal
is about 4 times smaller. M\'{e}ndez \etal emphasize how, while the star-galaxy 
separation software is only reliable to $V_\AB\approx 27.3$, misclassification
at fainter magnitudes turns more galaxies into stars than viceversa, as 
diffuse and faint features disappear first. The number of point-like sources 
detected in the range $27.3<V_\AB<29.8$ should then be considered as an 
absolute upper limit to the
number of stars (or quasars) observed in this magnitude interval. There are,
however, no obvious candidate high-$z$ QSOs in the 
published catalogues of unresolved objects in the HDF, as first remarked by 
Elson \etal In analogy with the case of star-forming galaxies, in fact,
Lyman continuum absorption arising because of the ubiquitous presence 
of \HI along the line of sight is so severe that faint quasars will drop out 
of the F450W image altogether by $z\approx 4$ (Paper I; Haiman, Madau, \&
Loeb 1998). At $z\gta 4.5$, line blanketing from the
\Lya forest will also cause a significant apparent depression ($\Delta
V_\AB\gta 1$ mag) in the F606W continuum image. High redshifts ($z>4$) quasars
are expected then to have very red $B-V$ and reddish $V-I$
colors. By contrast, all of the faint ($I_\AB<25.5$) stellar sources in the
M\'{e}ndez \etal (1996) sample have $(B-V)_\AB<0.6$ and $(V-I)_\AB<1.8$. 
A recent detailed search for faint QSOs in the HDF by Conti \etal
(1998) confirms the lack of a very red compact population: down to a 50\%
completeness limit of $V_\AB=29.6$ ($I_\AB=28.6$), no $z>4$ quasar candidates
are found. 

While a detailed comparison between deep imaging data and recent theoretical 
modeling of the evolution of the quasar LF has appeared elsewhere
(Haiman \etal 1998), it is important in this context to take note of the 
interesting constraints that the lack of red, unresolved objects in the HDF 
may pose on reionization scenarios which rely on a large number density of 
low-luminosity quasars at early epochs. Figure 3 shows the
predicted number-$I_\AB$ magnitude relation of active galaxies in the redshift 
range $4<z<5.5$ for the best-fit model discussed in the previous section and 
an illustrative variant, one that is characterized by a steeper LF with slope 
$\beta_1=2.0$  and a comoving space density that remains constant above 
$z=2.5$ instead of dropping. This alternative evolution model has been 
choosen in order to 
boost the actual emission rate of hydrogen-ionizing photons from QSOs at 
$z\sim 5$ by about a factor of 5. It is clear from Figure 3 that 
this model predicts too many red faint stellar sources in the field of view.
On the other hand, a large population of faint AGNs at high-$z$ would be 
consistent with the data if, at these faint magnitude levels and high image 
resolution, the host galaxies of active nuclei could actually be resolved.

We shall see in \S~3 below how, taken at face value, our results on the 
production of photons above 1 ryd by QSOs may pose some problems to the 
idea that a quasar-dominated background is responsible for maintaining the 
intergalactic gas in a highly ionized state at early epochs. Here we will 
simply note that the ability of quasars to overcome recombinations depends 
sensitively on the clumpiness of the medium. The emission rate of ionizing
photons in our standard QSO model, $\dot {\cal N}_Q\approx 4\times 10^{50}\,
\ndotunits$ at $z\approx 5$, is just a fraction of the total number of hydrogen 
recombinations in a clumpy, fast recombining universe, $R\approx 1.5\times 
10^{51} (\Omega_b h_{50}^2 /0.08)^2 C_{30}\,$ s$^{-1}$ Mpc$^{-3}$ if 
$T\approx 10^4$ K. 

\subsubsection{Star-forming Galaxies}

Galaxies with ongoing star-formation are another obvious source of Lyman
continuum photons. The tremendous progress in our understanding of faint galaxy
data made possible by the recent identification of star-forming galaxies at
$2\lta z\lta 4$ in ground-based surveys (Steidel \etal 1996a) and in the HDF
(Steidel \etal 1996b; Madau \etal 1996; Lowenthal \etal 1997) has provided new
clues to the long-standing issue of whether galaxies at high redshifts can 
provide a significant contribution to the ionizing background flux. Since the 
rest-frame UV continuum at 1500 \AA\ (redshifted into the visible band for a
source at $z\approx 3$) is dominated by the same short-lived, massive stars
which are responsible for the emission of photons shortward of the Lyman edge,
the needed conversion factor, about one ionizing photon every 10 photons at
1500 \AA, is fairly insensitive to the assumed IMF and is independent of the
galaxy history for $t\gg 10^{7.3}\,$ yr.\footnote{Shortward of the Lyman edge, 
however, the differences in the predicted ionizing radiation from model 
atmospheres of hot stars can actually be significant (see, e.g., Charlot 1996).}

A composite ultraviolet luminosity function of Lyman-break galaxies at 
$z\approx 3$ has been recently derived by Dickinson (1998). It is based
on spectroscopically (about 375 objects) and photometrically selected 
galaxies from the ground-based and HDF samples, and spans about a factor
50 in luminosity from the faint to the bright end. Because of the uncertanties
that still remain in the rescaling of the HDF data points to the ground-based 
data, the comoving luminosity density at 1500 \AA\ is estimated to vary 
within the range 1.6 to $3.5\times 10^{26}\emunits$. The ``best guess'' 
Schechter fit gives $\log\ (\phi_*/{\rm Gpc^{-3}})=6.1$, $\alpha=1.38$, and
$M_\AB^*(1500)=-20.95$ (Dickinson 1998), the magnitude at the ``break''
corresponding to a 
star-formation rate slightly in excess of $10\sfr$ (Salpeter IMF). Figure 4 
shows the Lyman continuum luminosity function of galaxies at $z\approx 3$ 
(at all ages $\gta 0.1$ Gyr one has $L(1500)/L(912)\approx 6$ for a Salpeter
mass function and constant star formation rate, Bruzual \& Charlot 1998), 
compared to the distribution of QSO luminosities at the same redshift. The 
comoving ionizing emissivity due to Lyman-break galaxies is $4.2\pm 1.5
\times 10^{25}\emunits$, between 2 and 4 times higher than the 
estimated quasar contribution at $z=3$. 

This number neglects any correction for dust extinction and intrinsic
\HI absorption. While it has been pointed out by many authors (e.g.,  
Meurer \etal 1997; Pettini \etal 1997; Dickinson 1998) that the colors of 
Lyman-break galaxies are redder than
expected in the case of dust-free star-forming objects, the prescription for a
``correct'' de-reddening is still unknown at present (note that redder spectra
may also results from an aging population or an IMF which is rather poor in
massive stars). A Salpeter IMF, $E_{\rm B-V}=0.1$ model with SMC-type dust in a
foreground screen, for example, has been found to reproduce quite well the
rest-frame ultraviolet colors of the HDF ``UV dropouts'' (Madau, Pozzetti, 
\& Dickinson 1998). In this model the color excess $E_{912-1500}=
1.64E_{\rm B-V}$ is rather small
and can be safely neglected in correcting from observed rest-frame far-UV to
the Lyman edge. However, for typical dust-to-gas ratios, it is the \HI
associated with dust that would greatly reduce the flux of Lyman continuum 
photons able to escape into the intergalactic space. The data point plotted 
in Figure 2 assumes a value of $f_{\rm esc}=0.5$ for the unknown fraction of 
ionizing photons which escapes the galaxy \HI layers into the intergalactic 
medium (e.g., Madau \& Shull 1996; Giallongo, Fontana, \& Madau 1997). The 
possible existence
of a numerous population of galaxies below the detection threshold, i.e. having
star formation rates $<0.5\sfr$, with a space density well in excess of
that predicted by extrapolating to faint magnitudes the $\alpha=1.38$ best-fit
Schechter function, will be discussed in \S~4.

The LF of Lyman-break galaxies at $z\gta 4$ is very uncertain. An analysis
of the $B$-band dropouts in the HDF -- candidate star-forming objects at
$3.5<z<4.5$ -- seems to imply a decrease in the comoving UV galaxy emissivity
by about a factor of 2 in the interval $2.75\lta z\lta 4$ (Madau \etal 1996, 
1998), but the error bars are still rather large. Adopting a $L(1500)$ 
to $L(912)$ conversion factor of 6, we estimate a comoving ionizing emissivity 
of $\sim 2 \times 10^{25} f_{\rm esc} \emunits$ at $z\approx 4$. One 
should note that, while a population of highly reddened galaxies at high 
redshifts would be missed by the dropout color technique (which isolates 
sources that have blue colors in the optical and a sharp drop in the 
rest-frame UV), it seems unlikely that very dusty objects with $f_{\rm esc}
\ll 1$, such as the recently discovered population of sub-mm sources (e.g. 
Barger \etal 1998)  would contribute in any significant manner to the ionizing 
metagalactic flux. 

\section{Reionization of the Universe}

In the previous section we have reviewed the basic theory of cosmological 
radiative transfer in a clumpy universe, demonstrated the local character of the
Lyman continuum flux at early epochs, and discussed the role of bright quasars 
and star-forming galaxies as candidate sources of photoionization at high 
redshifts. In this and the following sections we will try to shed some light 
on the time-dependent reionization process, and introduce the concept of a 
critical emission rate of hydrogen-ionizing photons per unit cosmological
volume.

In inhomogeneous reionization scenarios, the history of the transition from a
neutral IGM to one that is almost fully ionized can be statistically 
described by
the evolution with redshift of the {\it volume filling factor} or porosity
$Q(z)$ of \HII, \HeII, and \HeIII regions. The radiation emitted by spatially
clustered stellar-like and quasar-like sources -- the number densities and 
luminosities of which may change rapidly as a function of redshift -- 
coupled with
absorption processes in a medium with a time-varying clumping factor, all
determine the complex topology of neutral and ionized zones in the universe.
When $Q<<1$ and the radiation sources are randomly distributed, the ionized
regions are spatially isolated, every UV photon is absorbed somewhere in the
IGM, and the ionization process cannot be described as due to a statistically
homogeneous radiation field (Arons \& Wingert 1972; Meiksin \& Madau 1993). 
While the final size of the expanding ionized bubbles is only limited by the 
individual source lifetime, the timescale for full recombination is 
rather short and relict ionization zones may recombine faster than 
they can accumulate (cf Arons \& Wingert 1972). As $Q$ grows, the crossing of 
ionization fronts becomes more and more common, and the neutral phase shrinks 
in size until the reionization process is completed at the ``overlap'' epoch, 
when every point in space is exposed to Lyman continuum radiation and $Q=1$. 

\subsection{Time Evolution of Expanding Cosmological \bHII Regions}

When an isolated point source of ionizing radiation turns on, the ionized
volume initially grows in size at a rate fixed by the emission of UV photons,
and an ionization front separating the \HII and \HI regions propagates
into the neutral gas. Most photons travel freely in the ionized bubble, and are
absorbed in a transition layer. The evolution of an expanding \HII region is 
governed by the equation 
\begin{equation}
{dV_I\over dt}-3HV_I={\dot N_{\rm ion}\over \bar{n}_\nH}-{V_I\over 
\bar{t}_{\rm rec}}, \label{eq:dVdt} 
\end{equation}
(Shapiro \& Giroux 1987), where $V_I$ is the proper
volume of the ionized zone, $\dot N_{\rm ion}$ is
the number of ionizing photons emitted by the central source per unit time, and
all other symbols have their usual meaning. Across the I-front the degree of
ionization changes sharply on a distance of the order of the mean free path of
an ionizing photon. When $\bar{t}_{\rm rec}\ll t$, the growth of the 
\HII region is
slowed down by recombinations in the highly inhomogeneous IGM, and its evolution
can be decoupled from the expansion of the universe. Just like in the static
case, the ionized bubble will fill its time-varying Str\"omgren sphere
after a few recombination timescales,
\begin{equation}
V_I={\dot N_{\rm ion}\bar{t}_{\rm rec}\over \bar{n}_\nH} 
(1-e^{-t/\bar{t}_{\rm rec}}). \label{eq:V} 
\end{equation}
While the volume that is actually ionized depends on the luminosity of the 
central source, the time it takes to produce an ionization-bounded
region is only a function of $\bar{t}_{\rm rec}$. One should point out that the 
use of a volume-averaged clumping factor in the recombination timescale is 
only justified when the size of the \HII region is large compared to the 
scale of the clumping, so that the effect of many clumps (filaments) within 
the ionized volume can be averaged over. This will be a good approximation 
either at late epochs, when the IGM is less dense and \HII zones have had time 
to grow, or at earlier epochs if the ionized bubbles 
are produced by very luminous sources like quasars or the stars within halos 
collapsing from very high-$\sigma$ peaks. From equation
(\ref{eq:dis}), the mean free path between absorbers having neutral columns 
$>N_\nHI$ is $0.8\,$ Mpc h$_{50}^{-1}\,[(1+z)/6)]^{-4.5}\,(N_\nHI/10^{15}\,{\rm 
cm^{-2}})^{0.5}$: it is only on scales greater than this value that the
clumping can be averaged over. (Systems with $N_\nHI\ll 10^{14}$ cm$^{-2}$ do 
not dominate the mass, Miralda-Escud\`e \etal 1996; if the overdensity is an 
increasing function of $N_\nHI$, this implies, a fortiori, that low-$N_\nHI$
absorbers do not contribute significantly to the volume-averaged recombination 
rate.) On smaller scales underdense regions are 
ionized first, and only afterwards the UV photons start to gradually 
penetrate into the higher density gas.
 
With these caveats in mind, equation (\ref{eq:dVdt}) approximately holds for 
every isolated source of ionizing photons in the IGM. The filling factor of 
\HII regions in the 
universe, $Q_\nHII$, is then equal at any given instant $t$ to the integral
over cosmic time of the rate of ionizing photons emitted per hydrogen atom and
unit cosmological volume by all radiation sources present at earlier epochs, 
$\int_0^t \dot n_{\rm ion}(t')dt'/\bar{n}_\nH(t')$, {\it minus} the rate of 
radiative recombinations, $\int_0^t Q_\nHII(t')dt'/\bar{t}_{\rm rec}(t')$. 
Differentiating one gets
\begin{equation}
{dQ_\nHII\over dt}={\dot n_{\rm ion}\over \bar{n}_\nH}-{Q_\nHII\over 
\bar{t}_{\rm rec}}.  \label{eq:qdot}
\end{equation}
It is this simple differential equation -- and its equivalent for
expanding helium zones -- that statistically describes the transition
from a neutral universe to a fully ionized one, independently, for a given
emissivity, of the complex and possibly short-lived emission histories of
individual radiation sources, e.g., on whether their comoving space density is
constant or actually varies with cosmic time. In the case of a time-independent
clumping factor, equation (\ref{eq:qdot}) has formal solution 
\begin{equation}
Q_\nHII(t)=\int_0^t dt'\, {\dot n_{\rm ion}\over \bar{n}_\nH}\,
\exp\left(-{t'\over \bar{t}_{\rm rec}}+{t'^2\over \bar{t}_{\rm rec}t }\right),
\end{equation}
with $\bar{t}_{\rm rec}\propto t'^2$. Assuming a constant comoving 
emissivity, this can be rewritten as 
\begin{equation}
Q_\nHII(t)={\dot n_{\rm ion}t\over \bar{n}_\nH}\,\left[1-{t\over \bar{t}_{\rm 
rec}} \,e^{t/\bar{t}_{\rm rec}}\,E_1(t/\bar{t}_{\rm rec})\right],
\end{equation}
where $E_1(x)$ is the exponential integral function. Contrary to the 
static case, the \HII regions will always percolate in an expanding universe
with constant comoving emissivity and $t\rightarrow \infty$. At high 
redshifts, and for an IGM with $C\gg 1$, one can expand around $t$ to find 
\begin{equation}
Q_\nHII(t)\approx {\dot n_{\rm ion}\over \bar{n}_\nH}\bar{t}_{\rm rec}. 
\label{eq:qa}
\end{equation}
The porosity of ionized bubbles is then approximately given by the number
of ionizing photons emitted per hydrogen atom in one recombination time. In 
other words, because of hydrogen recombinations, only a fraction $\bar{t}_{\rm 
rec}/t$
($\sim$ a few per cent at $z=5$) of the photons emitted above 1 ryd is
actually used to ionize new IGM material. The universe is completely reionized
when $Q=1$, i.e. when 
\begin{equation}
\dot n_{\rm ion} \bar{t}_{\rm rec}=\bar{n}_\nH. 
\label{eq:qone}
\end{equation}
While this last expression has been derived assuming a constant
comoving ionizing emissivity and a time-independent clumping factor, it is also
valid in the case $\dot n_{\rm ion}$ and $C$ do not vary rapidly over a
timescale $\bar{t}_{\rm rec}$. Figure 5 shows the \HII filling factor 
resulting from numerical integration of equation (\ref{eq:qdot}), as a function 
of redshift for the QSO photoionization model described in \S~2.4, and a clumpy 
IGM with (constant) $C=1, 10, 20$ and 30. The transition from a neutral to a 
ionized universe takes place too late in this model, as late as $z=3.4$ for 
$C=30$, and never before $z=4.5$ even in the limiting case of a uniform medium. 
One should stress here the illustrative character of the curves depicted in 
Figure 5; in the real universe the clumpiness of the ionized hydrogen component 
will increase with cosmic time as more baryons collapse, and may also 
be a function of the UV photon production rate (see discussion below). 

\subsection{The Case of Centrally Concentrated Clumps}

The very simple model described in the previous section, one where 
individual \HII regions expand at a rate that depends on $\dot n_{\rm ion}$ 
and on a mean clumping factor assumed independent of the UV production rate, 
may not provide a good description of the late stages of reionization, when 
the fraction of baryons collapsed in virialized structures is non negligible, 
the filling factor of ionized material is close to unity, and a significant
fraction of all hydrogen recombinations (the rate of which is proportional 
to the square of the mass times the clumping factor of the recombining gas) may 
occur in the gas surrounding localized ``islands'' of neutral 
material with a small volume filling factor, e.g., the Lyman-limit absorbers. 

Consider, as an illustrative example, a fully ionized diffuse
IGM after complete overlapping has occurred, but with denser, partially neutral
spheres embedded in it, each characterized by an isothermal ($r^{-2}$) radial 
dependence of the gas density. Irradiated from the outside by the background  
radiation field generated by the ionizing sources, the neutral core of each 
clump will be delimited by a ``Str\"omgren surface'' with radius proportional to
$J^{-1/3}$, which will shrink as the UV metagalactic flux builds up with time. 
The cross section of the neutral cores, which determines the mean free 
path $\Delta l$ of ionizing photons, scales then as $J^{-2/3}$ (hence $\Delta 
l\propto J^{2/3}$). Since $J\propto$ UV emissivity $\times\, \Delta l$, in this 
simple model $J$ and $\Delta l$ will scale at a given epoch as the cube 
and square of the 
UV production rate at that epoch, respectively. On timescales where the 
number density of clumps stays approximately constant, the clumping factor 
scales as the inverse cube of the Str\"omgren radius,
increasing therefore as the cube of the photon production rate. 
Spheres with different density profiles show a change of behaviour when the 
radial power-law slope (gas density $\propto r^{\alpha}$, with $\alpha=-2$ in 
the case of isothermal spheres) becomes flatter than $\alpha=-3/2$. For 
these less peaked density distributions, the total 
recombination rate converges towards small radii: the Str\"omgren surface 
suddenly moves inward when the UV flux rises above some 
threshold, allowing the photon mean free path to increase rapidly. (In the 
limiting case $\alpha=-3/2$ the clumping factor becomes independent of the
emission rate of ionizing photons.) 

The equations put forward in the previous section provide a good description of
the reionization process only when the clumps are not centrally concentrated. 
In this case the recombination rate depends on the filling factor Q and 
saturates when most of the volume is ionized; $J$ increases steeply, by a 
large factor, when $Q$ becomes $\sim 1$ and the \HII regions overlap. 
In the presence of dense clumps with steep density gradients, however, 
recombinations occur mainly just outside the Str\"omgren surfaces 
(of course the high-density regions may be filaments rather
than spheres, but that does not change the picture), and there is a different
relationship between $J$ and $\dot n_{\rm ion}$ compared to that predicted 
in a model where the clumping does not depend on the ionizing emissivity. The 
impact of Lyman-limit absorbers with steep density profiles on the reionization 
histories depicted 
in Figure 5 will depend on their baryonic content as well as on their volume 
filling factor, both of which are only poorly known at the moment.  
     
\subsection{Delayed \bHeII Reionization}

Because of its higher ionization potential, the most abundant (by a factor 
$\sim 100$) absorbing ion in the universe is not \HI but \HeII. The 
importance of intergalactic helium in the context of this work stems from the 
possibility of detecting the effect of ``incomplete'' \HeII reionization in the 
spectra of $z\sim 3$ quasars as, depending on the clumpiness of the IGM (Madau 
\& Meiksin 1994) and on the spectrum of UV radiation sources (Miralda-Escud\`e 
\& Rees 1993), the photoionization of  singly ionized helium may be delayed 
until much later than for \HI.  

Since \HI and \HeI do not absorb a significant fraction of $h\nu>54.4\,$ eV
photons, the problem of \HeII reionization can be decoupled from that of 
other ionizations, and the equivalent of equation (\ref{eq:qdot}) for \HeIII 
expanding regions becomes
\begin{equation}
{dQ_\nHeIII\over dt}={\dot n_{\rm ion,4}\over \bar{n}_\nHe}-{Q_\nHeIII\over 
\bar{t}_{\rm rec,\nHeIII}}, \label{eq:qhe} 
\end{equation}
where $\dot n_{\rm ion,4}$ now takes into account only photons above 4 ryd, 
and $\bar{t}_{\rm rec,\nHeIII}$ is $\sim 6.5$ times shorter than the hydrogen 
recombination timescale if ionized hydrogen and doubly ionized helium have
similar clumping factors. It is interesting to note that, if the intrinsic 
spectrum of ionizing sources has slope $L(\nu)\propto \nu^{-1.8}$, the first 
terms on the right-hand side of equations (\ref{eq:qdot}) and (\ref{eq:qhe}) 
are actually equal, and a significant delay between the complete overlapping 
of \HII and \HeIII regions can only arise if recombinations are important.
This effect is illustrated in Figure 5, where the expected evolution of the 
\HeIII filling factor is plotted for our standard QSO-photoionization model: 
\HeII reionization is never completed before $z=3$ in models with $C\gta 10$. 
While a significant contribution to the UV background at 
1 ryd from massive stars may push the hydrogen reionization epoch to higher 
redshifts, the ratio between the number of \HeII and \HI Lyman continuum 
photons emitted from star-forming galaxies is only about two percent (Leitherer
\& Heckman 1995), five times smaller than in typical QSO spectra. It is 
likely then that recent observations of intergalactic \HeII absorption at high 
redshift may provide a rather direct test of quasar evolution models.

To date, a \HeII absorption trough has been detected in the spectra of four
distant QSOs. The {\it HST} and {\it Hopkins Ultraviolet Telescope} 
observations do not resolve individual absorption lines, but rather 
provide an average optical depth 
below 304 \AA\ in the rest-frame: $\tau_\nHeII=1.00\pm 0.07$ towards 
HS1700+6416 ($z_Q=2.74$, Davidsen, Kriss, \& Zheng 1996), $\tau_\nHeII>1.5$ 
towards PKS 1935-692 ($z_Q=3.18$, Jakobsen 1997), and $\tau_\nHeII>1.7$ 
towards Q0302-003 ($z_Q=3.29$, Jakobsen \etal 1994; Hogan, Anderson, \& 
Rugers 1996). These, together with the spectrum of HE2347-4342 ($z_Q=2.89$, 
Reimers \etal 1997), which reveals patchy absorption with low \HeII opacity
($\tau_\nHeII\approx 0.5$) ``voids'' alternating several Mpc sized regions 
with vanishing flux ($\tau_\nHeII=4.8^{+\infty}_{2.0}$), suggest that helium 
absorption does not increase smoothly with lookback time, but rather in 
the abrupt manner expected in the final stages of inhomogeneous \HeII       
reionization by quasar sources. A detailed assessment of the implications 
of late \HeII reionization, which may also have received some support from 
the recent observation of a sharp drop in the ratio of \SiIV to \CIV 
absorption in the \Lya forest at $z\approx 3$ (Songaila 1998), will be the 
subject of another paper. Here it is important to remark that, while delayed
\HeII reionization in a clumpy IGM appears to be naturally linked to the   
observed decline in the space density of quasars beyond $z\sim 3$, the 
complete overlapping of \HeIII regions occurs instead much earlier ($z\gg 5$) 
in models that predict many more faint QSOs at high redshifts (Haiman \& Loeb 
1998).   
    
\section{Discussion}

We have seen in the previous sections that, in the approximation the clumping 
can be averaged over, only the photons emitted within one recombination 
timescales can actually be used to ionize new material. As $\bar{t}_{\rm rec}
\ll t$ at high redshifts, it is possible to compute at any given epoch a 
critical value for the photon emission rate per unit cosmological comoving 
volume, $\dot {\cal N}_{\rm ion}$,
independently of the (unknown) previous emission history of the universe: only
rates above  this value will provide enough UV photons to ionize the IGM by 
that epoch. One can then compare our determinations of $\dot {\cal N}_{\rm
ion}$ to the estimated contribution from QSOs and star-forming galaxies. 
Equation (\ref{eq:qone}) can then be rewritten as
\begin{equation}
\dot {\cal N}_{\rm ion}(z)={\bar{n}_\nH(0)\over \bar{t}_{\rm rec}(z)}=(10^{51.2}\,
\ndotunits)\, C_{30} \left({1+z\over 6}\right)^{3}\left({\Omega_b 
h_{50}^2\over 0.08}\right)^2. 
\label{eq:caln}
\end{equation}
The uncertainty on this critical rate is difficult to estimate, as it depends 
on the clumping factor of the IGM (scaled in the expression above to the 
value inferred at $z=5$ from numerical simulations, Gnedin \& Ostriker 1997)
and the nucleosynthesis constrained baryon density. A quick exploration of the 
available parameter space indicates that the uncertainty on $\dot 
{\cal N}_{\rm ion}$ could easily be of order $\pm 0.2$ in the log. The 
evolution of the critical rate as a function of redshift is plotted in Figure 
2. While $\dot {\cal N}_{\rm ion}$ is comparable to the quasar contribution at 
$z\gta 3$, there is some indication of a significant deficit of Lyman 
continuum photons at $z=5$. For bright, massive galaxies to produce enough UV 
radiation at
$z=5$, their space density would have to be comparable to the one observed at
$z\approx 3$, with most ionizing photons being able to escape freely from the
regions of star formation into the IGM. This scenario may be in conflict with  
direct observations of local starbursts below
the Lyman limit showing that at most a few percent of the stellar ionizing
radiation produced by these luminous sources actually escapes into the IGM (Leitherer
\etal 1995).\footnote{Note that, at $z=3$, Lyman-break galaxies would radiate 
more ionizing photons than QSOs for $f_{\rm esc}\gta 30\%$.}~If, on the other 
hand, faint QSOs with (say) $M_\AB=-19$ at rest-frame ultraviolet frequencies 
were to provide {\it all} the required ionizing flux, their 
comoving space density would be such ($0.0015\,$Mpc$^{-3}$) that about 50 
of them would expected in the HDF down to $I_\AB=27.2$.  At $z\gta 5$, they 
would appear very red in $V-I$ as the \Lya forest is shifted into the visible.
This simple model can be ruled out, however, as there is only a handful (7) 
of sources in the HDF with $(V-I)_\AB>1.5$ mag down to this magnitude limit.
  
It is interesting to convert the derived value of $\dot {\cal N}_{\rm ion}$  
into a ``minimum'' star formation rate per unit (comoving) volume, $\dot 
\rho_*$ (hereafter we assume $\Omega_bh_{50}^2=0.08$ and $C=30$):
\begin{equation}
{\dot \rho_*}(z)=\dot {\cal N}_{\rm ion}(z) \times 10^{-53.1} f_{\rm esc}^{-1}
\approx 0.013 f_{\rm esc}^{-1} \left({1+z\over 6}\right)^3\ \sfrd. \label{eq:sfr} 
\end{equation}  
The conversion factor assumes a Salpeter IMF with solar metallicity, and has
been computed using Bruzual \& Charlot (1998) population synthesis code. 
It can be understood by noting that, for each 1 $M_\odot$ of stars formed, 
8\% goes into massive stars with $M>20 M_\odot$ that dominate the 
Lyman continuum luminosity of a stellar population. At the end of the C-burning
phase, roughly half of the initial mass is converted into helium and carbon,
with a mass fraction released as radiation of 0.007. About 25\% of the energy
radiated away goes
into ionizing photons of mean energy 20 eV. For each 1 $M_\odot$ of stars
formed every year, we then expect 
\begin{equation} 
{0.08\times 0.5 \times 0.007 \times 0.25\times M_\odot c^2\over 
20 {\,\rm eV\,}} {1\over  {\rm 1\, yr}} \sim 10^{53}\ndotun
\end{equation} 
to be emitted shortward of 1 ryd. Note that the star formation density given in
equation (\ref{eq:sfr}) is comparable with the value directly ``observed'' 
(i.e., uncorrected for dust reddening) at $z\approx 3$ (Madau \etal 1998; 
Dickinson 1998). If the volume emissivity has spectrum
$\epsilon(\nu)\propto \nu^{-\alpha_s}$, the ``minimum'' background
intensity at the Lyman limit is, from equations (\ref{eq:dll}),
(\ref{eq:nion}), and (\ref{eq:caln}) 
\begin{equation}
J_{912}(z)={1\over 4 \pi} \dot {\cal N}_{\rm ion}(z) h\alpha_s (1+z)^3\Delta
l(\nu_L)\approx 10^{-22} \, \alpha_s \left({1+z\over
6}\right)^{1.5}\, \uvunits.
\end{equation}

The same massive stars that dominate the
Lyman continuum flux also manufacture and return most of the metals to the 
ISM. In the approximation of instantaneous recycling, the rate of ejection of 
newly sinthesized heavy elements which is required
to keep the universe ionized at redshift $z$ is, from equation (\ref{eq:sfr}),
\begin{equation}
{\dot \rho_Z}(z)=y(1-R){\dot \rho_*}(z)\gta 3.5\times 10^{-4}
\left({y\over 2 Z_\odot}\right) \left({1+z\over 6}\right)^3 f_{\rm esc}^{-1}\, 
\sfrd, \label{eq:mfr}
\end{equation}
where $y$ is the net, IMF-averaged ``yield'' of returned metals, $Z_\odot=0.02$,
and $R\approx 0.3$ is the mass fraction of a generation of stars that is 
returned to the interstellar medium. At $z=5$, and over a timescale of $\Delta 
t=0.5\,$ Gyr corresponding to a formation redshift $z_f=10$, such a rate would 
generate a mean metallicity per baryon in 
the universe of
\begin{equation}
\langle Z \rangle\approx {8\pi G {\dot \rho_Z}(5) \Delta t\over 3 H_0^2 
\Omega_b}\gta 0.002 \left({y\over 2 Z_\odot}\right) f_{\rm esc}^{-1}Z_\odot,
\end{equation}
comparable with the level of enrichment observed in the \Lya forest at
$z\approx 3$ (Songaila 1997): more than 2\% of the present-day stars would need 
to have formed by $z\sim 5$. It has been suggested (e.g., Miralda-Escud\`e 
\& Rees 1998) that a large number of low-mass galactic halos, expected to
form at early times in hierarchical clustering models, might be responsible for 
photoionizing the IGM at these epochs. 
According to the spherically-symmetric simulations of
Thoul \& Weinberg (1996), photoionization heating by the UV background flux 
that builds up after the overlapping epoch completely suppresses the cooling 
and collapse of gas inside the shallow potential wells of halos 
with circular velocities $\lta 35\,\kms$. Halos with circular speed 
$v_{\rm circ}=50\,\kms$, corresponding in top-hat spherical collapse to a 
virial temperature $T_{\rm vir}=0.5\mu m_p v_{\rm circ}^2/k\approx 10^{5}\,$K 
and halo mass $M=0.1v_{\rm circ}^3/GH\approx 4\times 10^9 [(1+z)/6]^{-3/2}
h_{50}^{-1}\, \msun$, appear instead largely immune to this external feedback
(but see Navarro \& Steinmetz 1997). In these systems rapid cooling by 
atomic hydrogen can then take place and a significant fraction, 
$f\Omega_b$, of their total mass may be converted into stars over a timescale 
$\Delta t$ comparable to the Hubble time (with $f$ close 
to 1 if the efficiency of forming stars is high). If high-$z$ dwarfs with 
star formation rates $f\Omega_bM/\Delta t\sim 0.3 f_{0.5} \Delta t_{0.5}^{-1}\, 
\sfr$  were actually responsible for keeping the universe ionized at $z\sim 
5$, their comoving space density would have to be 
\begin{equation}
{0.013\, f_{\rm esc}^{-1}\sfrd\over 0.6 f {\, \rm M_\odot yr^{-1}}}
\sim 0.1 \left({f_{\rm esc} f\over 0.25}\right)^{-1} {\,\rm Mpc^{-3}},
\end{equation}
two hundred times larger than the space density of present-day galaxies 
brighter than $L^*(4400)$, and about five hundred times larger than that of 
Lyman-break objects at $z\approx 3$ with $M<M^*_\AB(1500)$,
i.e. with star formation rates in excess of $10\,\sfr$. Only a rather
steep luminosity function, with Schechter slope $\alpha\sim 2$, would
be consistent with such a large space density of faint dwarfs and, at the same
time, with the paucity of brighter $B$- and $V$-band dropouts observed in the
HDF. The number density on the sky would be $\approx 0.2\,$ arcsec$^{-2}$, 
corresponding to more than three thousands sources in the HDF. With a typical 
apparent magnitude at $z=5$ of $I_\AB \sim 29.5$ mag 
(assuming $f=0.5$), these might be too faint to 
be detected by the {\it HST}, but within the range of the proposed {\it Next
Generation Space Telescope} (Stockman \etal 1998). 

These estimates are in reasonable agreement (although towards the low side) 
with the calculations of Miralda-Escud\`e \& Rees (1998), who used the observed
metallicity of the \Lya forest as the starting point of their investigation. 
A higher density of sources -- which would therefore have to originate from 
lower amplitude peaks -- would be required if the typical efficiency of star 
formation and/or the escape fraction of ionizing photons were low, $(f, f_{\rm 
esc})\ll 1$. In this case the dwarfs could still be detectable  
if a small fraction of the gas turned into stars in very short bursts
(there would then be an extra parameter associated with their duty cycle, 
in addition to $f_{\rm esc}$ and $f$). A reduction of 
the star formation rate in halos with low circular velocities (necessary 
in hierarchical cosmogonies to prevent too many baryons from turning into stars
as soon as the first levels of the hierarchy collapse, White \& Frenk 1991)
may result from the heating and possible expulsion of the gas due to 
repeated supernova (SN) explosions after only a very small number of stars 
have formed. Recent numerical simulations (Mac Low \& Ferrara 1998) show, 
however, that metal-enriched material from SN ejecta is accelerated to 
velocities larger than the escape speed from such systems far more easily than 
the ambient gas. If the same population of dwarf galaxies that may keep the 
universe ionized at $z\approx 5$ were also responsible for polluting 
the IGM with heavy elements, it is interesting to ask whether these atoms 
could diffuse uniformly enough to account for the observations of weak but
measurable \CIV absorption lines associated with the \Lya forest clouds 
(Lu \etal 1998; Songaila 1997). Our fiducial estimate of 0.1 sources 
per comoving Mpc$^3$ implies that, in order to pollute the entire IGM, the 
metal-enriched material would have to be expelled to typical distances of order
1 Mpc, which would require an ejection speed of about $200[(1+z)/6]^{1/2}\,
\kms$. In fact, the heavy elements may be restricted to filaments within only 
(say) 20 proper kpc (100-150 comoving kpc) of the halos, and in this case 
it would be enough to accelerate the outflowing metals to velocities comparable 
to the escape velocity, rather than to the higher speeds associated with 
SN ejecta. The required diffusion of heavy elements would be a more serious  
constraint if the relevant galaxies were rarer and more luminous, as would 
happen if they originated from 3-$\sigma$ peaks and star formation was 
postulated to occur with higher efficiency than in more typical peaks.

\acknowledgments
We have benefited from useful discussions with A. Conti, M. Haehnelt, J. 
Miralda-Escud\'e, and especially D. Sciama. Support for this work was provided 
by NASA through ATP grant NAG5-4236 (PM), and by the Royal Society (MJR). 

\references

Arons, J., \& Wingert, D.~W. 1972, \apj, 177, 1

Barger, A. J., Cowie, L. L., Sanders, D. B., Fulton, E., Taniguchi, Y.,
Sato, Y., Kawara, K., Okuda, H. 1998, Nature, 394, 248

Bechtold, J. 1994, ApJS, 91, 1

Bechtold, J., Weymann, R.~J., Lin, Z., \& Malkan, M. 1987, \apj, 315, 180

Boyle, B.~J., Shanks, T., \& Peterson, B.~A. 1988, \mnras, 235, 935

Bruzual, A. G., \& Charlot, S. 1998, in preparation
 
Cen, R., Miralda-Escud\'e, J., Ostriker, J.~P., \& Rauch, M. 1994, \apj, 437, L9

Charlot, S. 1996, in ASP Conf. Ser. 98, From Stars to Galaxies: The Impact of
Stellar Physics on Galaxy Evolution, ed. C. Leitherer, U. Fritze-von 
Alvensleben, \& J. Huchra (San Francisco: ASP), 275

Cheng, F.-Z., Danese, L., De Zotti, G., \& Franceschini, A. 1985, \mnras, 
212, 857

Ciardi, B., \& Ferrara, A. 1997, \apj, 483, L5

Conti, A., Kennefick, J. D., Martini, P., \& Osmer, P. S. 1998, submitted
to AJ (astro-ph/9808020)

Copi, C. J., Schramm, D. N., \& Turner, M. S. 1994, Science, 267, 192

Couchman, H.~M.~P., \& Rees, M.~J. 1986, \mnras, 221, 53

Cowie, L. L., Songaila, A., Kim, T.-S., \& Hu, E. M. 1995, \aj, 109, 1522

Davidsen, A. F., Kriss, G. A., \& Zheng, W. 1996, Nature, 380, 47

Dickinson, M. E. 1998, in The Hubble Deep Field, ed. M. Livio, S. M. Fall, \&
P. Madau (Cambridge: Cambridge University Press), in press (astro-ph/9802064) 

Elson, R. A. W., Santiago, B. X., \& Gilmore, G. F. 1996, New Astronomy, 
1, 1

Fall, S.~M., \& Pei, Y.~C. 1993, \apj, 402, 479

Flynn, C., Gould, A., \& Bahcall, J. N. 1996, \apj, 466, L55

Francis, P.~J., Hewett, P.~C., Foltz, C.~B., Chaffee, F.~H., Weymann,
R.~J., \& Morris, S.~L. 1991, \apj, 373, 465

Fukugita, M., \& Kawasaki, M. 1994, \mnras, 269, 563

Giallongo, E., D'Odorico, S., Fontana, A., McMahon, R. G., Savaglio, S.,
Cristiani, S., Molaro, P., \& Trevese, D. 1994, \apj, 425, L1

Giallongo, E., Fontana, A., \& Madau, P. 1997, \mnras, 289, 629

Giroux, M. L., \& Shapiro, P. R. 1996, \apjs, 102, 191

Gnedin, N. Y., \& Ostriker, J. P. 1997, \apj, 486, 581

Gunn, J.~E., \& Peterson, B.~A. 1965, \apj, 142, 1633

Haardt, F., \& Madau, P. 1996, \apj, 461, 20 (Paper II)

Haehnelt, M. G., Natarajan, P., \& Rees, M. J. 1998, MNRAS, in press

Haehnelt, M. G., \& Rees, M. J. 1993, \mnras, 263, 168

Haiman, Z., \& Loeb, A. 1997, \apj, 483, 21

--------- 1998, \apj, in press

Haiman, Z., Madau, P., \& Loeb, A. 1998, submitted to ApJ (astro-ph/9805258)

Hartwick, F.~D.~A., \& Schade, D. 1990, ARA\&A, 28, 437

Hawkins, M. R. S., \& Veron, P. 1996, \mnras, 281, 348

Hernquist, L., Katz, N., Weinberg, D.~H., \&  Miralda-Escud\'e, J.
1996, \apj, 457, L51

Hogan, C. J., Anderson, S. F., \& Rugers, M. H. 1997, \aj, 113, 1495

Hook, I. M., Shaver, P. A., \& McMahon, R. G. 1998, in The Young Universe:
Galaxy Formation and Evolution at Intermediate and High Redshift, ed. S.
D'Odorico, A. Fontana, \& E. Giallongo (San Francisco: ASP), in press
(astro-ph/9801213)

Hu, E.~M., Kim,  T.-S., Cowie, L. L., Songaila, A., \& Rauch, M. 1995, \aj,
110, 1526

Jakobsen, P. 1997, in Structure and Evolution of the Intergalactic Medium 
from QSO Absorption Line Systems, ed. P. Petitjean \& S. Charlot (Paris: 
Nouvelles Fronti\`eres), 57

Jakobsen, P., Boksenberg, A., Deharveng, J. M., Greenfield, P., 
Jedrzejewski, R., \&  Paresce, F. 1994, Nature, 370, 35 

Kennefick, J. D., Djorgovski, S. G., \& de Carvalho, R. R. 1995, \aj, 110, 2553
(KDC) 

Kim, T.-S., Hu, E. M., Cowie, L. L., \& Songaila, A. 1997, \aj, 114, 1 

Lanzetta, K. M., Wolfe, A. M., \& Turnshek, D. A. 1995, \apj, 440, 435

Laor, A., Fiore, F., Elvis, M., Wilkes, B. J., \& McDowell, J. C. 1997,
\apj, 477, 93

Leitherer, C., Ferguson, H.~C., Heckman, T.~M., \& Lowenthal,
J.~D. 1995, \apj, 454, L19

Leitherer, C., \& Heckman, T.~M. 1995, \apjs, 96, 9

Lowenthal, J.~D., Koo, D.~C., Guzman, R., Gallego, J., Phillips, A.~C., Faber,
S.~M., Vogt, N.~P., Illingworth, G.~D., \& Gronwall, C. 1997, \apj, 481, 673 

Lu, L., Sargent, W. L. W., Barlow, T. A., \& Rauch, M. 1998, submitted to AJ
(astro-ph/9801189)

Madau, P. 1992, \apj, 389, L1

--------- 1995, \apj, 441, 18 (Paper I)

Madau, P., Ferguson, H.~C., Dickinson, M.~E., Giavalisco, M., 
Steidel, C.~C., \& Fruchter, A. 1996, \mnras, 283, 1388

Madau, P., \& Meiksin, A. 1994, \apjl, 433, L53

Madau, P., Meiksin, A., \& Rees, M. J. 1997, \apj, 475, 429

Madau, P., Pozzetti, L., \& Dickinson, M. E. 1998, \apj, 498, 106

Madau, P., \& Shull, J.~M. 1996, \apj, 457, 551

Magorrian, J., \etal 1997, submitted to AJ (astro-ph/9708072)

Mac Low, M.-M., \& Ferrara, A. 1998, submitted to ApJ (astro-ph/9801237) 

Meiksin, A., \& Madau, P. 1993, \apj, 412, 34

M\'{e}ndez, R. A., Minniti, D., De Marchi, G., Baker, A., \& Couch, W. J.
1996, \mnras, 283, 666

Meurer, G.~R., Heckman, T.~M., Lehnert, M.~D., Leitherer, C., \& Lowenthal, J.
1997, \aj, 114, 54

Miralda-Escud\'e, J., Cen, R., Ostriker, J.~P., \& Rauch, M. 1996, \apj, 471, 
582

Miralda-Escud\'e, J., \& Ostriker, J.~P. 1990, \apj, 350, 1

Miralda-Escud\'e, J., \& Rees, M. J. 1993, \mnras, 260, 617

--------- 1998, \apj, 497, 21 

Navarro, J. F., \& Steinmetz, M. 1997, \apj, 478, 13

Osmer, P. S. 1982, \apj, 253, 280

Ostriker, J. P., \& Gnedin, N. Y. 1996, \apj, 472, L63

Ostriker, J. P., \& Heisler, J. 1984, \apj, 278, 1

Paresce, F., McKee, C., \& Bowyer, S. 1980, \apj, 240, 387

Peebles, P.~J.~E. 1993, Principles of Physical Cosmology
(Princeton: Princeton University Press)

Pei, Y. C. 1995, \apj, 438, 623

Petitjean, P., Webb, J. K., Rauch, M., Carswell, R. F., \& Lanzetta, K. M.
1993, \mnras, 262, 499

Pettini, M., Steidel, C. C., Dickinson, M. E., Kellogg, M., Giavalisco, M.,
\& Adelberger, K. L. 1998, in The Ultraviolet Universe at Low and High
Redshift, ed. W. Waller, (Woodbury: AIP Press), in press (astro-ph/9707200)
 
Press, W. H., \& Rybicki, G. B. 1993, \apj, 418, 585

Reimers, D., K\"{o}hler, S., Wisotzki, L., Groote, D., Rodriguez-Pascual,
P., \& Wamsteker, W. 1997, A\&A, 327, 890

Sargent, W.~L.~W., Steidel, C.~C., \& Boksenberg, A. 1989, \apjs, 69, 703

Schneider, D.~P., Schmidt, M., \& Gunn, J.~E. 1991, \aj, 101, 2004

Schmidt, M., Schneider, D.~P., \& Gunn, J.~E. 1995, \aj, 110, 68

Sciama, D. W. 1995, \apj, 448, 667

Shapiro, P. R., \& Giroux, M. L. 1987, \apj, 321, L107

Shaver, P. A., Wall, J. V., Kellerman, K. I., Jackson, C. A., \&
Hawkins, M. R. S. 1996, Nature, 384, 439

Songaila, A. 1997, \apj, 490, L1

---------  1998, \aj, in press (astro-ph/9803010)

Songaila, A., Cowie, L. L., \& Lilly, S. J. 1990, \apj, 348, 371

Steidel, C.~C., Giavalisco, M., Dickinson, M.~E., \& Adelberger, K. 1996b, \aj,
112, 352 

Steidel, C.~C., Giavalisco, M., Pettini, M., Dickinson, 
M.~E., \& Adelberger, K. 1996a, \apj, 462, L17

Stengler-Larrea, E., \etal 1995, \apj, 444, 64

Stockman, H. S., Stiavelli, M., Im, M., \& Mather, J. C. 1998, in  ASP
Conf. Ser. 133, Science with the Next Generation Space Telescope, ed. E. 
Smith \& A. Koratkar (San Francisco: ASP), 24


Tegmark, M., Silk, J., \& Blanchard, A. 1994, \apj, 420, 484

Thoul, A. A., \& Weinberg, D. H. 1996, \apj, 465, 608

Tytler, D. 1987, \apj, 321, 49

Warren, S.~J., Hewett, P.~C., \& Osmer, P.~S. 1994, \apj, 421, 412 (WHO)


White, S. D. M.,\& Frenk, C. S. 1991, \apj, 379, 25

Wright, E. L. 1990, \apj, 353, 411

Zhang, Y., Anninos, P., \& Norman, M. L. 1995, \apj, 453, L57

Zhang, Y., Meiksin, A., Anninos, P., \& Norman, M. L. 1998, \apj, 495, 63

Zheng, W., Kriss, G. A., Telfer, R. C., Crimes, J. P., \& Davidsen, A. F.
1998, \apj, 492, 855


\vfill\eject
\begin{figure}
\plotone{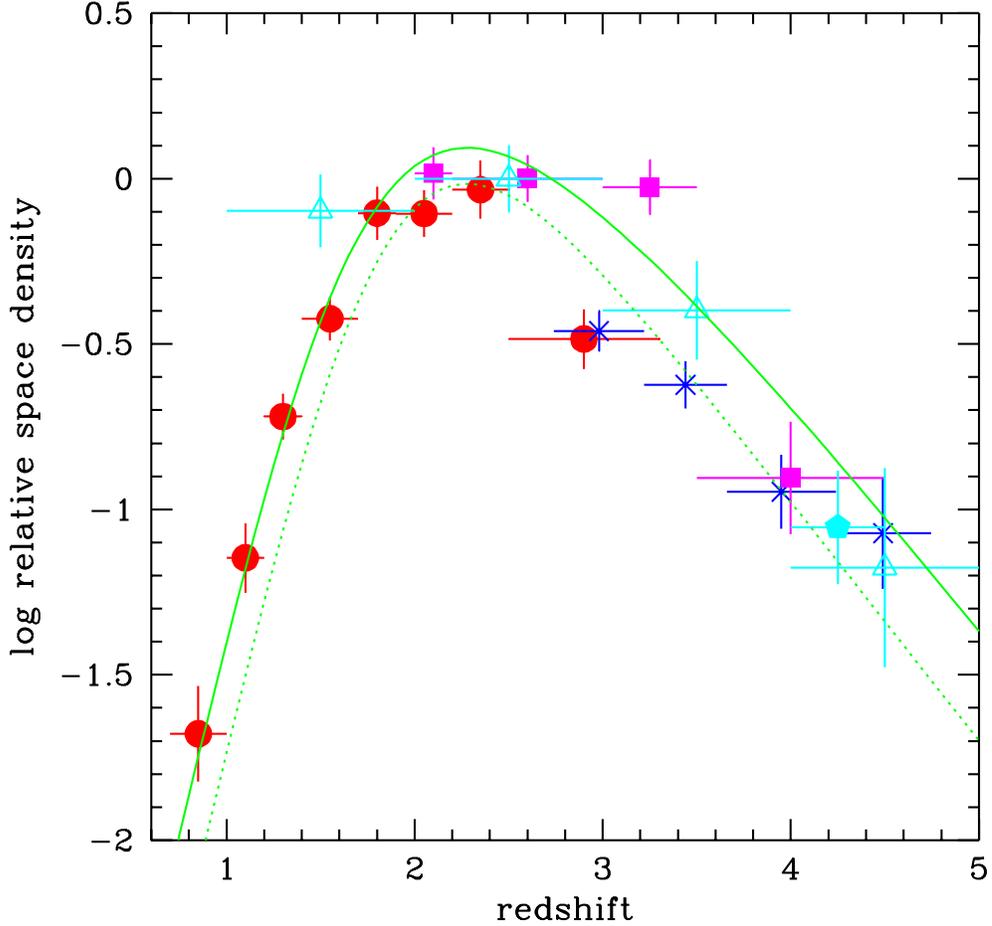}
\caption{Comoving space density of bright QSOs as a function of redshift for
$(h_{50}, q_0, \alpha_s)=(1.0, 0.5, 0.5)$. The data points with error bars are
taken from Hartwick \& Shade (1990) {\it (filled dots)}, WHO {\it (filled 
squares)}, Schmidt \etal (1995) {\it (crosses)}, and KDC {\it (filled
pentagon)}. The points have been normalized to the $z=2.5$ space density of
quasars with $M_B<M_{\rm lim}=-26$ ($M_B<M_{\rm lim}=-27$ in the case of KDC) 
as given by WHO. The curves ({\it solid line} for 
$M_{\rm lim}=-26$, and {\it dotted line} for $M_{\rm lim}=-27$) have been
computed from eqs. (\ref{eq:phi}), (\ref{eq:lz}), and (\ref{eq:NQ}), with
$\log\ (\phi_*/{\rm Gpc^{-3}})=2.95$, $\beta_1=1.64$, $\beta_2=3.52$,
$M_*(0)=-22.35$, $z_*=1.9$, $\zeta=2.58$, and $\xi=3.16$, and have been
normalized to WHO as well. The {\it empty
triangles} show the space density (normalized to the peak) of the Parkes
flat-spectrum radio-loud quasars with $P>7.2\times 10^{26}\,$ W Hz$^{-1}$
sr$^{-1}$ (Hook \etal 1998). 
\label{fig1}}
\end{figure}

\begin{figure}
\plotone{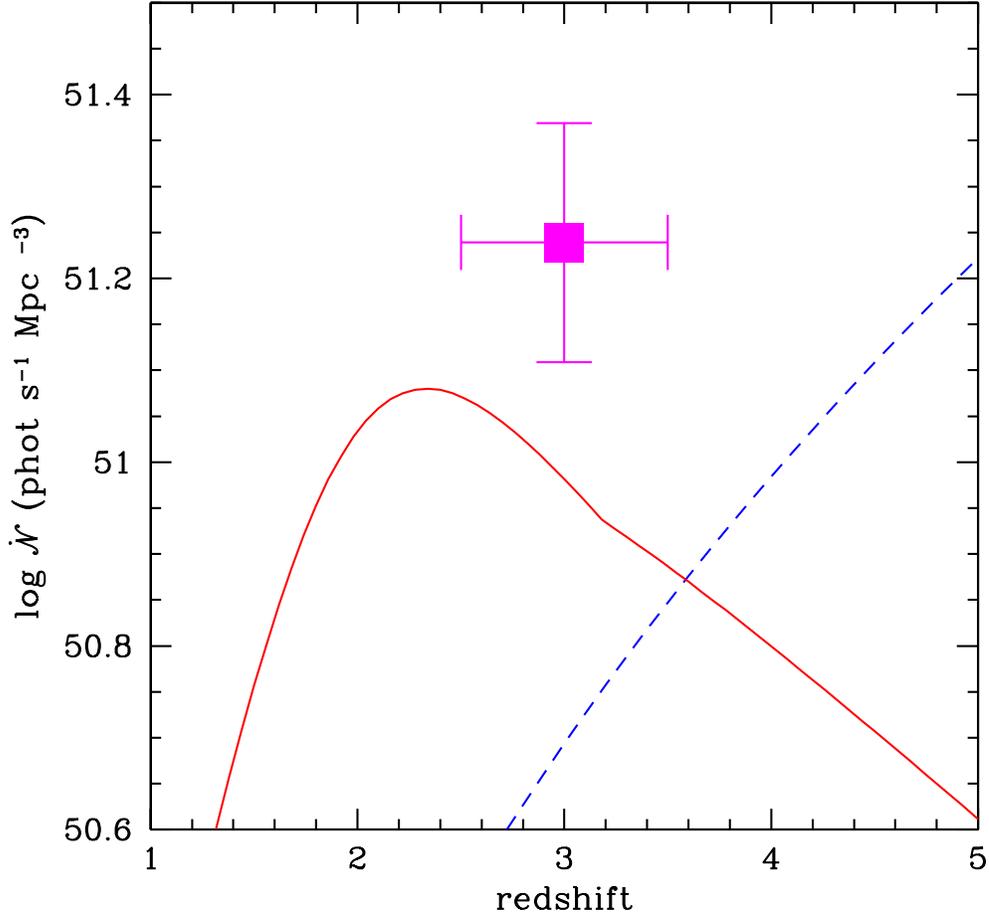}
\caption{Comoving emission rate of hydrogen Lyman continuum photons ({\it solid
line}) from QSOs, compared with the minimum rate ({\it dashed line}) which is
needed to fully ionize a fast recombining (with clumping factor $C=30$) 
Einstein--de Sitter universe with 
$h_{50}=1$ and $\Omega_b=0.08$. Models based on photoionization by quasar 
sources appear to fall short at $z=5$. See text for details on the assumed
quasar luminosity function and spectral energy distribution.  The data point
with error bars show the estimated contribution of star-forming galaxies 
at $z\approx 3$. The fraction 
of Lyman continuum photons which escapes the galaxy \HI layers into the 
intergalactic medium is taken to be $f_{\rm esc}=0.5$.
\label{fig2}} 
\end{figure}

\begin{figure}
\plotone{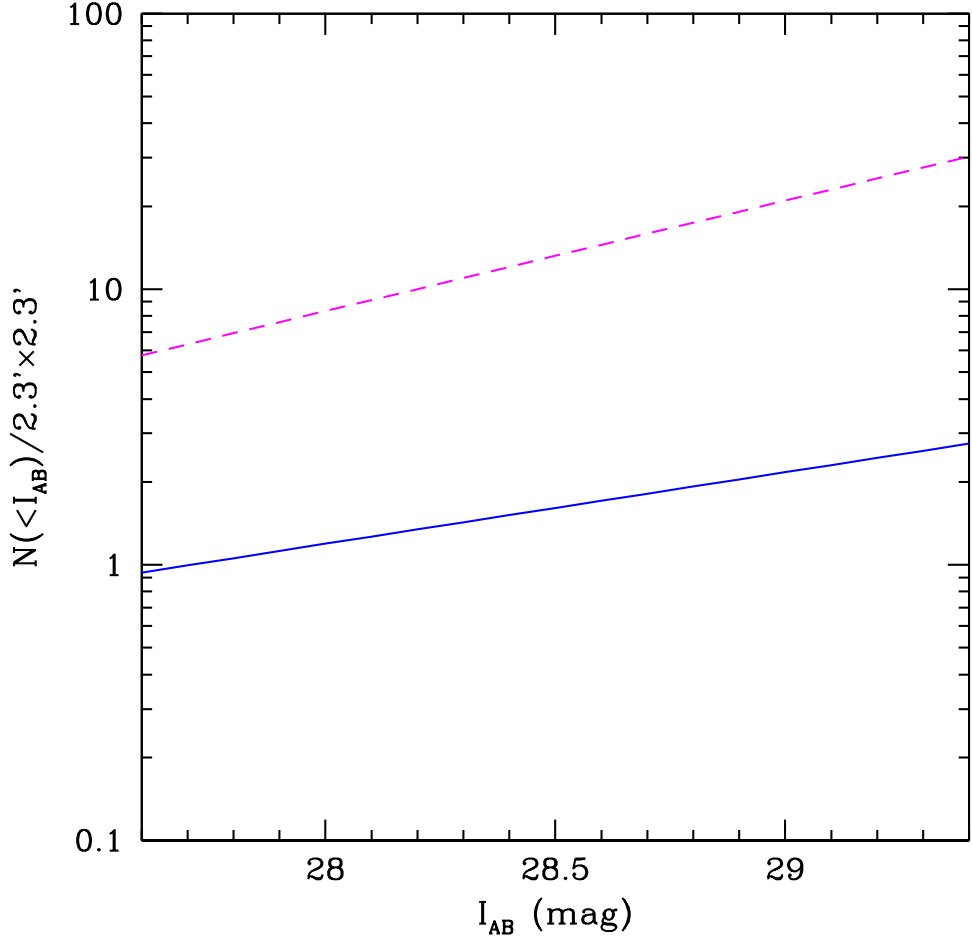}
\caption{Cumulative number-magnitude relation of quasars in the redshift range
$4.0<z<5.5$. The {\it solid line} shows the prediction for our standard model,
one in which the faint end of the QSO luminosity function has slope
$\beta_1=1.64$ and tracks the turnover observed in the space density of bright
quasars at $z\gta 3$. The {\it dashed line} shows the increased number of
sources expected in the case of a steeper, $\beta_1=2.0$, luminosity function
and a comoving space density which stays constant above $z=2.5$. The latter 
evolution scenario provides, within the errors, enough UV photons to keep the 
universe ionized at $z\approx 5$, but might be inconsistent with the lack 
of red, faint stellar objects observed in the {\it Hubble Deep Field}. 
Note that, in the assumed cosmology, $I_\AB=28$ corresponds to an absolute 
magnitude of $M_\AB(1360\,$\AA)$=-18.2$ at $z=5$, at the very faint end of the
AGN luminosity function. 
\label{fig3}}
\end{figure}

\begin{figure}
\plotone{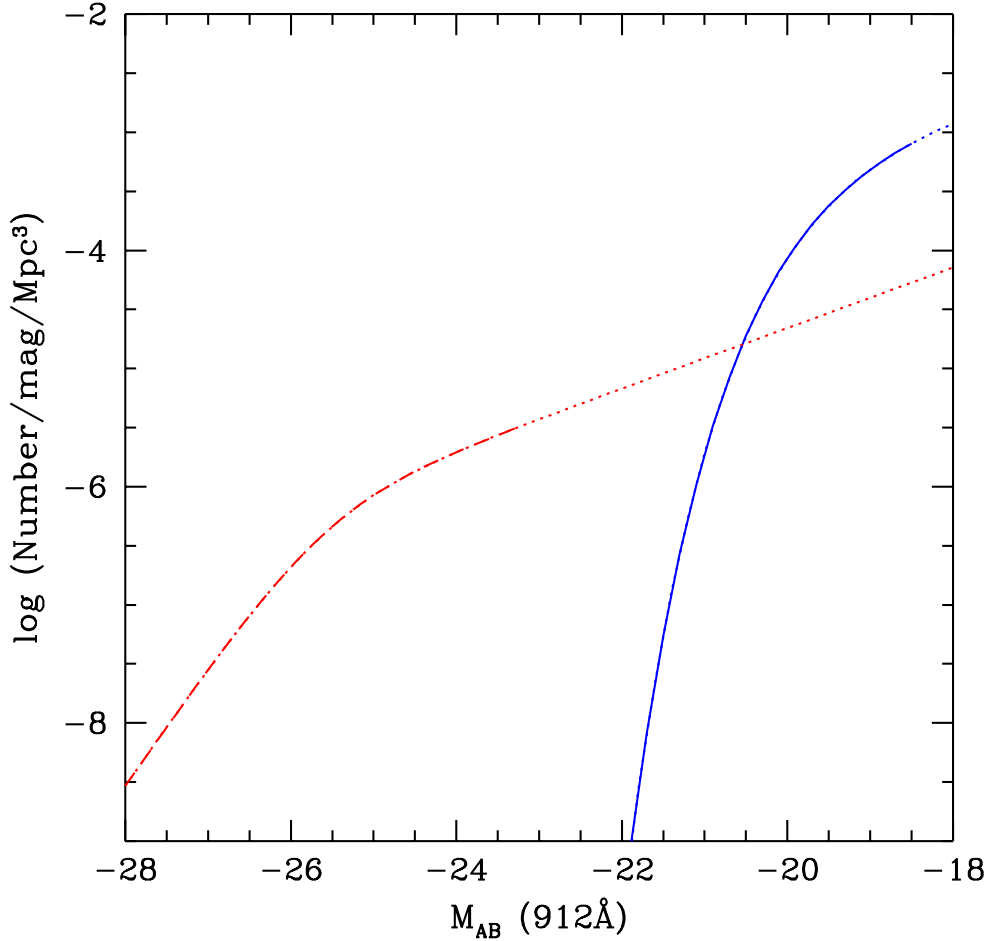}
\caption{The $912\,$\AA\ luminosity function of galaxies at $z\approx 3$ ({\it
solid line}), compared to the distribution of QSO luminosities at the same
redshift ({\it dashed line}). The latter has been derived assuming a spectral
slope of $\alpha_s=0.5$. The former assumes a Salpeter IMF with constant 
constant star formation rate (age=1 Gyr): $M_\AB(912\,$\AA)$=-19$ corresponds 
to a rate of $13\sfr$. The solid and dashed lines represent functional 
fits to the data points, and the dotted lines their extrapolation. 
\label{fig4}} 
\end{figure}

\begin{figure}
\plotone{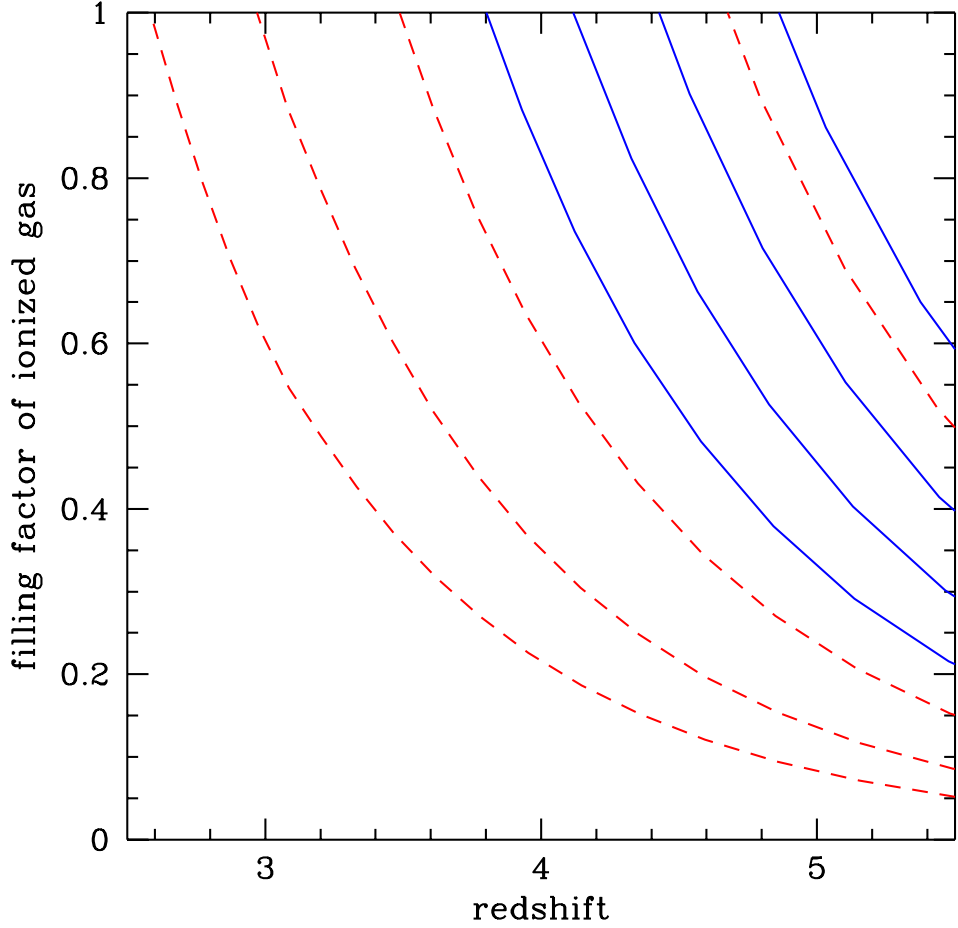}
\caption{The evolution of the \HII ({\it solid lines}) and \HeIII ({\it
dashed lines)} filling factors as a  function of redshift for an inhomogeneous 
universe where photoionization is dominated by QSOs, and clumps with steep
density gradients  make only a negligible contribution to the total 
recombination rate. The IGM density is taken to be 
$\Omega_b h_{50}^2=0.08$, and, from right to left, the four curves assume
a time-independent clumping factor of $C=1, 10, 20,$ and 30 ($C=1$ in the case
of a uniform IGM). The QSO intrinsic 
spectrum varies as $\nu^{-1.8}$ shortward of the hydrogen Lyman edge. Note 
how the ionization of \HI (\HeII) is never completed before $z=4.0$ ($z=3.0$) 
in models with $C\ge 10$.
\label{fig5}}
\end{figure}

\end{document}